\documentclass{article}

 \usepackage[preprint,nonatbib]{neurips_2024}

\usepackage[utf8]{inputenc} 
\usepackage[T1]{fontenc}    
\usepackage{hyperref}       
\usepackage{url}            
\usepackage{booktabs}       
\usepackage{amsfonts}       
\usepackage{nicefrac}       
\usepackage{microtype}      
\usepackage{xcolor}         
\usepackage{tikz}
\usetikzlibrary{shapes.geometric, positioning}

\usepackage[sorting=none]{biblatex}
\bibliography{hyperrefs}

\usepackage{tcolorbox}
\tcbuselibrary{listingsutf8} 

\title{AI Safety Frameworks Should Include Procedure for Model Access Decisions}

\author{Edward Kembery \\
  ERA Fellow \\
  \texttt{edwardkembery@outlook.com} 
  \And
  Tom Reed \\
  ERA Fellow}

\begin{document}

\maketitle

\begin{abstract}
   
   The downstream use cases, benefits, and risks of AI models depend significantly on what sort of access is provided to the model, and who it is provided to. Though existing safety frameworks and AI developer usage policies recognise that the risk posed by a given model depends on the level of access provided to a given audience, the procedures they use to make decisions about model access are ad hoc, opaque, and lacking in empirical substantiation. This paper consequently proposes that frontier AI companies build on existing safety frameworks by outlining transparent procedures for making decisions about model access, which we term Responsible Access Policies (RAPs). We recommend that, at a minimum, RAPs should include the following: i) processes for empirically evaluating model capabilities given different styles of access, ii) processes for assessing the risk profiles of different categories of user, and iii) clear and robust pre-commitments regarding when to grant or revoke specific types of access for particular groups under specified conditions.
\end{abstract}

\newpage
\section{Executive Summary} 

Existing frontier AI safety frameworks recognise that the societal impact of an AI model depends on what style of access to the model is provided to which group(s). Some, such as Anthropic's, reference ``access controls'' explicitly \cite{anthropicResponsibleScalingUpdate}. However, current safety frameworks provide no procedures to ensure that these decisions about who should have what style of access are made responsibly.

Frontier AI companies should expand existing safety frameworks to account for access considerations. We refer to this body of procedure as \textit{Responsible Access Policies} (RAPs). They should set out \textit{general commitments} to follow best practices in governing model access, and \textit{specified procedure} for assessing the risks and benefits of providing different access styles to different groups:  
\vspace{-1mm} 

\begin{figure}[h]
    \centering
    \includegraphics[width=1\linewidth]{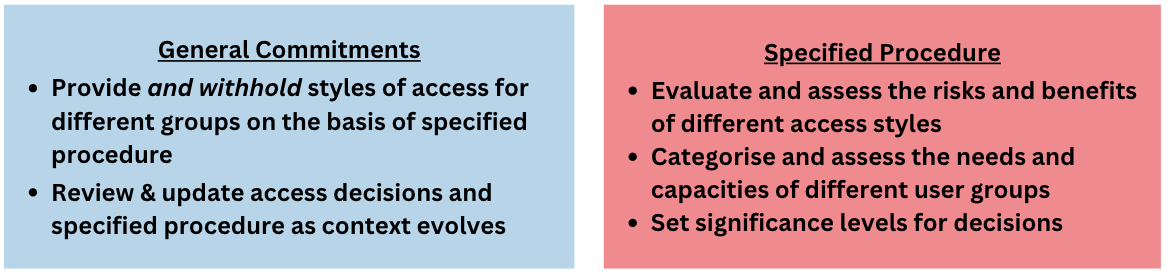}
\end{figure}
\vspace{-2mm} 

We stress that these decision-making processes should be clear and transparent to stakeholders. To make these decisions legible, we suggest that frontier AI companies use an \textit{Access Assessment Matrix}:
\vspace{-2mm} 


\begin{figure}[h]
    \centering
    \includegraphics[width=1\linewidth]{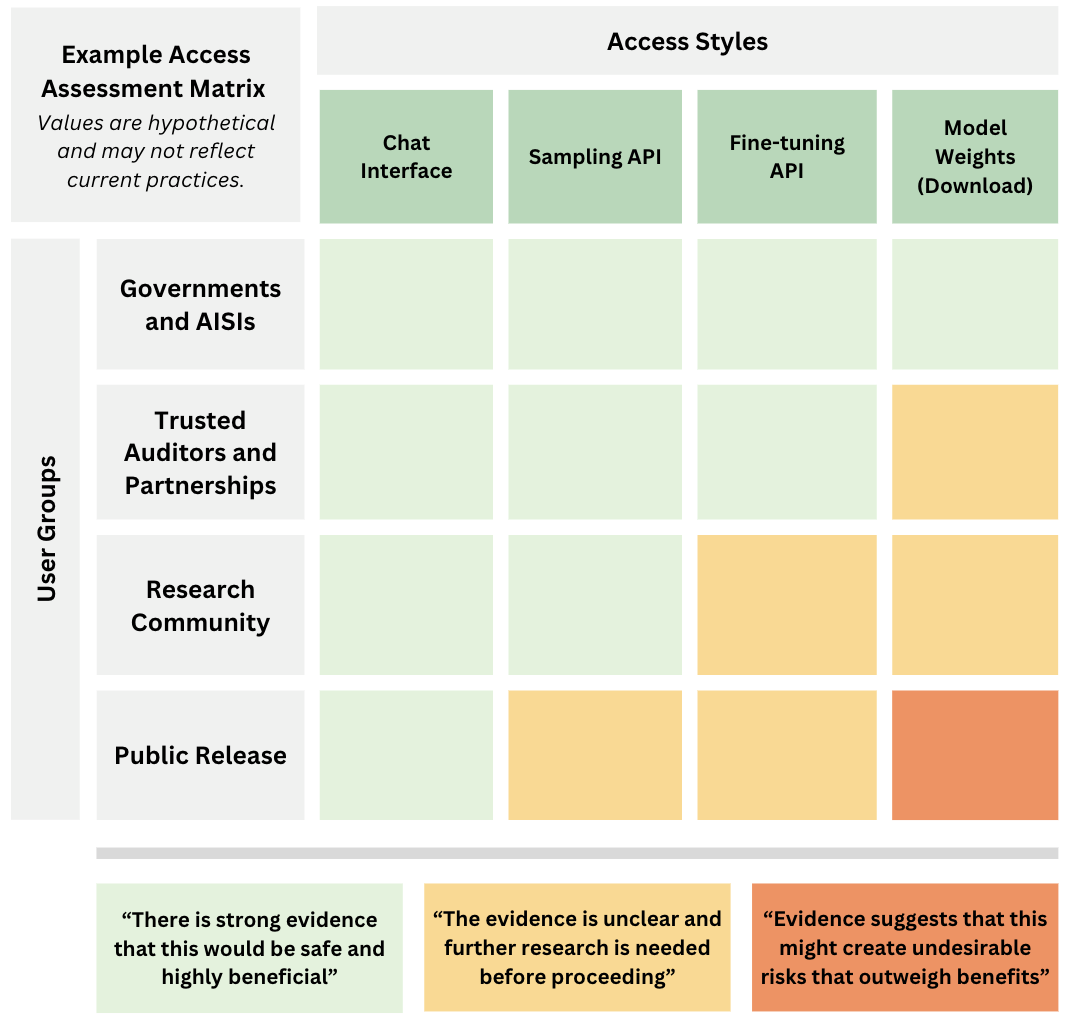}
\end{figure}

\newpage

\section{Introduction}

The rapid development of general-purpose AI has led several frontier developers to implement safety frameworks which help to structure decisions about how to minimise societal harm whilst promoting economic growth and respect for diverse cultural norms \cite{rsp, OpenAIPreparedness, deepmindIntroducingFrontier}. One purpose of these frameworks is to help guide decision-making about \textit{how} and \textit{to whom} developers release their models to \cite{ModelAccessGovernance, seger2023opensourcinghighlycapablefoundation, eiras2024near}. For instance, a recent update to Anthropic's RSP explicitly mentions that ``access controls'' and a ``tiered access system'' should be developed as part of commitments to exist safety frameworks \cite{anthropicResponsibleScalingUpdate}.

Incorporating access policy into safety frameworks is appropriate. General-purpose AI models can be made available to end users in a wide number of different ways which enable distinct uses cases, each of which may engender different downstream impacts. If access to information or components of a system is provided to users incautiously, it might pose societal risks by increasing the likelihood that powerful AI systems are misused \cite{seger2023opensourcinghighlycapablefoundation, davidevan, kapoor2024societalimpactopenfoundation} or stolen \cite{carlini2024stealingproductionlanguagemodel}. Providing clear frameworks for governing decisions would also create certainty for researchers and businesses that rely on publicly available models.

However, existing safety frameworks do not go far enough to support responsible model access governance. Decisions about model access at frontier AI companies are often ad hoc, opaque, and lacking in empirical support.\footnote{For instance, Meta's CEO stated that if ``there's some qualitative change in what the thing is capable of, and we feel like it's not responsible to open source it, then we won't'' \cite{dwarkeshpatelMarkZuckerberg}. Meta has not released any statements as to how these decisions would be made in practice.} Although Anthropic's recent updates are encouraging, no other frontier AI safety frameworks outline protocols for making informed decisions about how and to whom a model is deployed \cite{OpenAIPreparedness, deepmindIntroducingFrontier}.

Consequently, we contend that frontier AI companies should build on existing safety frameworks by including access considerations. First, we make the case for Responsible Access Policies (RAPs) as part of safety frameworks, building on model access research. Second, we set out a framework for RAPs, suggesting that these should incorporate i) empirical evaluation of access styles, ii) processes for classifying different categories of user and iii) clear pre-commitments on providing or retracting access to specified groups in defined circumstances. We emphasize that frontier AI companies have a real opportunity to set an example here{\textemdash}both for other AI companies and for regulators{\textemdash}helping to develop and implement responsible, empirically-driven model access governance. 

\subsection{Related Literature}

\textbf{Model Deployment Governance}. Frontier AI companies that have produced safety frameworks around AI systems include Anthropic \cite{rsp}, Google DeepMind \cite{deepmindIntroducingFrontier} and OpenAI \cite{OpenAIPreparedness}. However, these extent to which these frameworks are comprehensive or feasible is unclear \cite{alaga2024gradingrubricaisafety}, and other than Anthropic's recent update, they do not make explicit commitments to procedures for making decisions about which groups to provide access to under which circumstances. Comapnies like Meta, Mozilla and LAION \cite{fbOpenSource, NTIARequestforcomment} have associated themselves with `open-source' access policies, but may continue to do so as AI models develop. 

\textbf{Structured access and model access}. Structured transparency{\textemdash}a framework for information sharing designed to facilitate collaboration and information sharing whilst addressing privacy concerns \cite{trask2024privacytradeoffsstructuredtransparency}{\textemdash} has been applied to model access decisions around AI systems for several years \cite{shevlane2022structured}. Recent work has explored what styles of model access are advantageous for valuable safety research \cite{benbucknallsafety} and audits  \cite{longpre2024safe, mokander2023auditing, mokander2022conformity, mokander2021ethics, mokander2023operationalising}. Technical work has explored the extent to which it is possible to make model access governance robust: for instance, by assessing how easy it is to ``steal'' the weights of a production language model through interactions \cite{tramèr2016stealingmachinelearningmodels, carlini2024stealing}.

\textbf{Open vs closed-source debate}. Early work related to model access governance transferred a framing of `openness' from earlier software research (e.g. Millar et al., 2022 \cite{millar2022vulnerabilitydetectionopensource}). In practice, key intuitions have proved unsuitable for AI systems \cite{basdevant2024frameworkopennessfoundationmodels}, and there is as yet no official definition of `open source AI' \cite{OSIOpensource}. Some research has argued that we should move `beyond the open vs closed debate' \cite{carnegieendowmentBeyondOpen} towards `hybrid' access regimes that involve providing access to some aspects of the model but not others \cite{solaiman2023gradientgenerativeairelease}. Nonetheless, several papers have usefully focused on models with widely available model weights in order to identify theoretical marginal risks \cite{stanford2023open} and possible alternatives to open-weight deployment \cite{seger2023opensourcinghighlycapablefoundation}. Others have made a vigorous case for various definitions of open-source \cite{eiras2024near, NTIARequestforcomment}.

\subsection{Definitions}

\textbf{Model Aspects:} AI models are complex structures composed of various elements, such as code and weights \cite{basdevant2024frameworkopennessfoundationmodels}. They are the basis for AI systems, which involve additional infrastructure like GPUs and user interfaces \cite{basdevant2024frameworkopennessfoundationmodels, EUAIAct}. Organizations developing AI models must decide not only which components of the model to make accessible to different groups but also how to share broader information about the model, such as training data or sample outputs. We refer to this set of information as the \textit{model aspects}. 

\textbf{Access Styles:} When a developer makes an aspect of a model available to a user in a certain way (for instance, with certain permissions or restrictions on how the information is to be used), we refer to this as an \textit{access style}. Different access styles might include chat, fine-tuning, weights inspection, or weights modification  \cite{benbucknallsafety}. 

\textbf{Access Regimes:} When organizations develop an AI model, they may offer different access styles to various groups. For example, a company might provide access to the weights for an AI Safety Institute, API-based fine-tuning for trusted researchers, and a chat interface for the public. We refer to this collective set of access styles granted to all users as an \textit{access regime}. Different access regimes might create different use cases en masse, leading to different downstream impacts. 

\newpage

\section{The Case for Responsible Access Policies}

\subsection{Misgoverning model access could have serious societal consequences}

Incautious access policies could create greater societal risks \cite{davidevan, seger2023opensourcinghighlycapablefoundation}. Whilst there is already substantial literature investigating the potential of powerful AI systems to cause serious harm, \cite{anderljung2023frontierairegulationmanaging, weidinger2021ethical}, miscalibrated access governance might make these risks significantly worse, for three main reasons. First, it could make it easier to bypass the safeguards of models that would otherwise be robust to jailbreaking \cite{qi2023finetuningalignedlanguagemodels}, either by allowing detuning or making it easier to design custom jailbreaks. Second, unsecured models could spread globally, with no way to retract them, expanding both the geographic risk zone and the time frame of the risk \cite{seger2023opensourcinghighlycapablefoundation}. Thirdly, miscalibrated access could make it harder to have oversight on how models are used, reducing decision-makers' insight into evolving risks, thereby making them less prepared to address these issues. 

On the other hand, overly restrictive access policies might incur high opportunity costs \cite{eiras2024near}. Broadly speaking, these costs fall into three categories. First, providing insufficient access to safety organisations could delay the development of safety research \cite{benbucknallsafety, casper2024blackboxaccessinsufficientrigorous}, thus resulting in greater future risks from AI systems. Second, miscalibrated access might lead to underutilisation of valuable use cases, hampering the ability of the public to develop novel AI applications without substantially increasing risk \cite{fbOpenSource}. Third, overly restrictive access could lead to a more unequal distribution of decision-making power over advanced AI, perhaps disproportionately affecting users from particular socioeconomic backgrounds \cite{pipatanakul2023typhoonthailargelanguage, blasi2021systematic}. 

\subsection{Frontier AI companies might not necessarily make good decisions by default}

The current absence of frameworks for model access governance makes it difficult for developers to make good decisions about how and to whom they release their models. There are at least three reasons this is the case.    First, decision-makers are currently very uncertain about key issues, such as the relative increase in risk of harm caused by releasing a particular access modality to a particular audience. Empirical evidence as to potential risks and benefits of different access regimes is generally lacking \cite{carnegieendowmentBeyondOpen}. 

Second, current safety frameworks fall short in governing model access. While they outline how companies manage risks in developing and deploying frontier AI systems \cite{alaga2024gradingrubricaisafety}, they lack procedures for decisions on model access. For example, many of them do not address how the `risk category' of a model may vary by access style \cite{OpenAIPreparedness, deepmindIntroducingFrontier}. Nor do they clarify how user categories are classified or how these relate to threat models informing risk thresholds for deployment. OpenAI’s Preparedness Framework, for instance, mentions restricting dangerous models to trusted parties but lacks criteria for defining a trusted party \cite{OpenAIPreparedness}. (Furthermore, if the risk category of models was to depend significantly on what style of access was provided to particular parties, conventional safety frameworks may prove insufficient for reducing risks---however, this is beyond the scope of this paper \cite{alaga2024gradingrubricaisafety}).

Finally, organisational incentives might not be naturally aligned with responsible model access governance. For instance, pursuing one type of access regime might increase the number of customers willing to pay for a model and so be desirable for the company, even if this regime creates an uplift in the capabilities of malicious actors. 

\subsection{Governments and societies need more clarity about how this will develop in future}

Even if miscalibrated access regimes posed no marginal risk of harms, or if most frontier AI companies were already well-placed to make good decisions, there are still at least three reasons why developers should adopt RAPs. First, governments and industry that rely on AI models are likely to require clear guidance on the sort of model access they can expect to have in the future. Second, these policies should be transparent so as to encourage governments and researchers to audit them, improve them, and contest disagreements in a open manner. Third, whilst some frontier AI companies might be well situated to govern model access effectively, sharing best practises between companies, and around the world, might provide incentives towards developing good model access governance practises across industry.  

\newpage

\section{Incorporating Responsible Access Policies}

\subsection{Deployers should pre-commit to responsible model access governance processes}

Model governance pre-commitments, as implemented in safety frameworks, offer at least three key advantages for model access governance. They increase the likelihood of compliance, inform governments about the effectiveness of self-regulation, and provide clarity for those who depend on particular model access styles for their everyday use cases. RAPs should involve two types of pre-commitments: general commitments about the procedures that will be followed related to model access governance, and specified procedures related to access style `roll-forward' and `roll-back' decisions:

\textbf{General Commitments.} Frontier AI companies should provide general commitments as to how they intend to govern model access. This could include clauses to i) pre-commit not to release any access styles without a clear safety case to do so, ii) pre-commit to evaluating different access styles for frontier models in the context of different user groups, and iii) pre-commit to reviewing and potentially updating these decisions (whether for or against a particular access regime) on an ongoing basis as use-cases and technologies change.

\textbf{Specified Procedure.} Frontier AI companies should offer explicit commitments regarding decisions about whether to `roll-forward' or `roll-back' different access styles. These commitments may include: i) detailed protocols for conducting evaluations, including defined significance levels; ii) procedures for assessing the different use cases of groups concerned and how they might be affected by the decision; and iii) governance frameworks that demonstrate how this information will be integrated into decision-making processes, as well as identifying those groups accountable for the resulting decisions.

We suggest that frontier AI companies approach these decisions with caution, especially in regard to situations where providing access might mean that companies are not able to retract it in future (for instance, if a group was able to download model weights). In some cases, it may be appropriate to investigate mitigating strategies for reducing the risks associated with model access regimes (for instance, providing more robust watermarking for models with open-source access regimes), or reducing the burden on users of retracted access styles. 

\subsection{Decisions about access should be informed by empirical evaluations of the use cases and risks across diverse access styles and end-user categories}

The impact profile of a model is related both to its capabilities and the particular access regime it is subject to --- that is, both \textit{what} sort of access is provided and \textit{whom} they provided to. In the \textit{Access Assessment Matrix}, we refer to these as `styles of access' and `user groups' respectively. RAPs should incorporate evaluations that account for both before considering whether to provide particular styles of access to particular groups:

\textbf{Evaluating Access Styles}. Researchers already assess how different access styles might uplift different use cases for different groups \cite{benbucknallsafety}. Researchers should continue to explore how particular capabilities change between access styles such as a chat-based interface, secure fine-tuning access, and open weights release. They might also explore how different access styles might change in future (for instance, as synthetic data for fine-tuning models becomes better at eliciting greater capabilities from smaller models). Key priorities might be investigating this change across domains like Chem/Bio, Cyber, and Autonomous Replication, but researchers should also investigate the extent to which different styles of access might make it easier to bypass model safeguards and enable harmful use cases related to CSAM, NCII and disinformation. These evaluations might be supplemented by research into the more general long-term downstream effects of particular access styles by third-party organisations. 
 
\textbf{Accounting for User Profiles}. Frontier AI companies already assess how different capabilities might differently enable different groups \cite{openai2024bio}. Groups that frontier AI companies might want to provide access to include general research organisations, specified trusted AI evaluation organisations like AISIs, governments, and the public. These groups might have different goals, resources, time, and levels of skill at eliciting capabilities from models given particular access styles. Frontier AI companies should have reasonable views as to which groups have what level of skill, time and resources, and should, if necessary, assess how these are changing on an ongoing basis (for instance, as prompt-engineering skills improve or sophisticated scaffolds proliferate online). These user profiles should be paired with evaluations of the ranging capabilities afforded by different access styles (as described above), in order to inform decisions about whether providing a given access modality to a given category of end-user is high-risk, or alternatively, societally desirable. If evidence is unclear, it may be useful to incorporate a phase in which a model is tested for a sustained period in a secure and closely monitored API to understand how the model is used by a group.  

We expect that modelling different malicious actors using new technologies over different time frames with different resources will be a significant challenge. Frontier AI companies should collaborate with different actors in government to get better assessments of these as models become more powerful, and refrain from making definitive statements in lieu of sparse evidence or inconclusive results. We suggest that frontier AI companies should be especially wary of dangerous capabilities when considering irreversible access styles such as downloadable weights release \cite{davidevan}.  
    
\subsection{Decision-making procedures should be clear and transparent}

If RAPs are implemented but not operationalized effectively by frontier AI companies, they might diminish appetite for legislation without ensuring that model access is governed responsibly. If they are not transparent to the public and key stakeholders, they might be less useful for coordinating governance, increasing the possibility of unnecessary legislation, and creating uncertainty for groups who depend on particular access styles. Policies should therefore be clear and transparent:

\textbf{Clarity}. Making RAPs clear and operationalizable will require precise definitions which avoid unfairness, prevent confusion, and eliminate the potential for the creation of legal loopholes that could be exploited. It will be important to define access styles carefully, paying special attention to the extent to which it might be possible to elicit information about some parts of the models by accessing others \cite{carlini2024stealingproductionlanguagemodel}. It will also be important to define key terms such as `research organisations', `trusted third-parties' and `public' carefully so that these decisions can be adhered to and evaluated \cite{ModelAccessGovernance}. Definitions should be robust enough to provide external watchdogs clear signals when access procedures are not being properly followed, and to hold decision-makers accountable.

\textbf{Transparency}. Making RAPs transparent will require documentation on both general principles and specified procedures at various levels of detail. Appropriate documentation for processes might include i) a longer-form access policy document circulated internally detailing general commitments and specified procedure, ii) a medium-length summary of both aspects published online and iii) a short memo on critical issues tailored for government decision-makers. Appropriate reporting related to a specific decision to release a model for a particular access regime (say, to release fine-tuning access to the public) might include a long-form report detailing the evaluations procedure undertaken, the risks and benefits of the proposed access regime, possible implications of the regime, and the set of circumstances which would have to be met for the decision to change. It might also include a shorter document with key headlines for public release, in the style of a model card. The \textit{Access Assessment Matrix} (see Executive Summary) might be a useful way to build on existing data representation techniques used by safety frameworks to represent some of this information concretely. 

\section{Conclusion}

Whilst frontier AI companies have taken several important and necessary steps to ensure that they govern model deployment responsibly, model access governance procedures are still nascent, and unlikely to be sufficient to avoid societal risks and opportunity costs as models become more capable. Incorporating Responsible Access Policies into existing safety frameworks effectively will be crucial for ensuring that frontier AI companies make responsible decisions about model access. 

Doing so will have challenges. The different use cases afforded by different access styles may be unclear and change substantially over time as people find new ways to use the system, or technologies develop. New modalities like agents may affect the extent to which different access styles enable different use cases, requiring access regime changes to minimise risks. Specific decisions, such as retracting styles of access, may be difficult to implement and harmful to particular user groups if not executed with care. However, these challenges only serve to emphasise the importance of forethought, research, responsible pre-commitments and clear decision-making processes that could be supported by RAPs.

\newpage

\section{Acknowledgements}

This research was supported by the ERA Fellowship. The authors would like to thank the ERA Fellowship for its financial and intellectual support, and Ben Bucknall, Guarav Sett and Michael Chen in particular for their support and suggestions. 

\section{Errata}

An earlier version of this paper mistakenly included style components indicating that this paper had been accepted to the Socially Responsible Language Modelling Research (SoLaR) Workshop at NeurIPS 2024. This was not the case. We apologise. 

\section{Bibliography}

\printbibliography

\end{document}